# Quality of explanation of xAI from the prespective of Italian end-users: Italian version of System Causability Scale (SCS)

Carmine Attanasio [1[0000 0002 8294 4421]] , Alireza Mortezapour [1[0000-0001-6356-2244]]

[1] University of Salerno, Salerno, Italy
amortezapoursoufiani@unisa.it

**Abstract:**

**Background and aim:** Considering the scope of the application of artificial intelligence beyond the field of computer science, one of the concerns of researchers is to provide quality explanations about the functioning of algorithms based on artificial intelligence and the data extracted from it. The purpose of the present study is to validate the Italian version of system causability scale (I-SCS) to measure the quality of explanations provided in a xAI.

**Method:** For this purpose, the English version, initially provided in 2020 in coordination with the main developer, was utilized. The forward-backward translation method was applied to ensure accuracy. Finally, these nine steps were completed by calculating the content validity index/ratio and conducting cognitive interviews with representative end users.

**Results:** The original version of the questionnaire consisted of 10 questions. However, based on the obtained indexes (CVR below 0.49), one question (Question 8) was entirely removed. After completing the aforementioned steps, the Italian version contained 9 questions. The representative sample of Italian end users fully comprehended the meaning and content of the questions in the Italian version.

**Conclusion:** The Italian version obtained in this study can be used in future research studies as well as in the field by xAI developers. This tool can be used to measure the quality of explanations provided for an xAI system in Italian culture.

**Keyword:** Usability, Cause-effect, comprehensibility, explainability, user perception

# Introduction:

Today, various applications of artificial intelligence outside the boundaries of computer science are well proven [1]. Applications in the fields of medicine [2], agriculture [3], education [4], etc. have been well discussed and claimed. Precisely in this situation, we should know that the end users of data and processing based on artificial intelligence are not necessarily computer experts and other experts from different fields can potentially use this information [5]. So presenting a solutions for helping experts from other fileds to use data based on AI is necessary [6].

One of the solutions presented in this field is the introduction of xAI [7]. In this case, the data generated by artificial intelligence must be perceptible, interpretable and usable by the end-users alone [8]. That is, an expert user who works outside the field of computer science (for example, medicine) must be able to receive usable and perceptible concepts from the data obtained from artificial intelligence processes and use them in his/her specialized field. In this situation we can tell that end-users of AI data are considered in the loop [9, 10].

Attention to end users in computer science and human-computer interaction is not only limited to this case. For more than 40 years, with similar concepts, the end users as consumers of products have been of interest of researchers [11, 12]. If we want to look for the closest similar concept, we can refer to a concept called usability of a system [13]. In the concept of usability, the final product must be efficient, effective and satisfy the end user [14]. This can be considered in line with the concept called xAI, in which the end users of data based on artificial intelligence are considered important [15, 16]. Therefore, the ultimate goal of xAI can be summarized as providing a usable algorithm or AI data. It is in this situation that the end user can expect the effectiveness and efficiency from a AI data and at the end be satisfied with its use [17].

According to these two close concepts, in recent years, one of the usability criteria in xAI is the quality of the explanation provided for the AI data and algorithms [18]. To measure the quality of the explanations provided for an artificial intelligence algorithm, various methods have been developed according to the end user [16]. One of them, which was developed precisely according to one of the widely-accepted usability evaluation method, is system causability scale (SCS) [19].

As mentioned, this evaluation method, which was developed precisely according to the system usability scale (SUS, since 1986), was presented for the first time in 2020 [20]. Just like SUS, this new method (SCS) also has 10 questions. As mentioned, the ultimate goal of both tools is to consider the end-user's opinion about a product. Based on a definition presented by the main developer, SCS measures how useful explanations of xAI are and how usable the explanation interface for end-user is [19].

Usability assessment tools should be used according to the cultural backgrounds of end-users [21, 22]. Considering that the English version of the SCS has been presented, but the Italian version that can be used in the culture of Italy is not yet available, the purpose of this study is to validate the Italian version of this tool.

## Method:

The aim of this study was to develope the Italian version of the SCS. So xAI developers can quieckly analysis the quality of their provided explanations based on the opinion of their customers (the end users of their xAI product). The linguistic validation technique is used in the next 9 steps [23].

### 2.1. Giving permision from the main developer

In order to ensure the intellectual rights of the developer of this tool, it was allowed to translate it into Italian by email.

### 2.2. First translation from English to Italian

In this step, two bilingual expert (Native in Italian as well as fluent in English) independentely worked on the translation the questionnaire from English into Italian. The experts were contacted and shown each of them the questionnaire consisting of 10 questions in English and they were asked to translate them.

### 2.3. Comparison of the two translations

In this step, we examined each question/item and compared them looking for inconsistencies. The objective of this analysis was to verify the possible inconsistency and inaccuracy of the translations or to identify potential errors or ambiguities in the translation process.

### 2.4. Fix the inconsistency

In this passage the inconsistencies that have been found have been resolved thanks to counsult with both traslators and the mediator role of the first author. In general, the goal is to provide a consistent and accurate translation. We have thus obtained a single consistent translation.

**2.5. Backward translation from Italian to English**

This step required that the questionnaire be translated again from Italian to English by a native-speaking expert, without the latter seeing the original questionnaire. This expert is different from previous two experts and the main team of the research.

**2.6. Send the translation to the main developer**

This step involved sending the English version obtained in the previous step to the main author, who highlighted and corrected any errors so that the version is identical to the original.

**2.7. Initial Italian Version**

After considering some minor point from the main deveoper, we obtained the initial Italian version of the questionnaire. This version was identical to the main english version but needs some more steps. Here we present the initial Italian translation of the questionnaire which has 10 questions:

1. Ho trovato che i dati includevano tutti i fattori causali noti rilevanti con sufficiente precisione e granularità.
2. Ho compreso le spiegazioni nel contesto del mio lavoro.
3. Ho potuto modificare il livello di dettaglio su richiesta
4. Non ho avuto bisogno di supporto per capire le spiegazioni
5. Ho sentito che le spiegazioni mi hanno aiutato a capire la causalità.
6. Sono stato in grado di utilizzare le spiegazioni grazie alle mie conoscenze di base
7. Non ho trovato incongruenze tra le spiegazioni
8. Penso che la maggior parte delle persone imparerebbe a capire le spiegazioni molto velocemente.
9. Non ho avuto bisogno di ulteriori riferimenti nelle spiegazioni: ad es, linee guida mediche, regolamenti.
10. Ho ricevuto spiegazioni in modo tempestivo ed efficiente.

## 2.8. Evaluation of the validity of the questionnaire

Lawshi's method was used for analyzing the Content Validity Ratio (CVR) and Content Validity Index (CVI) [24]. Questionnaires were prepared and distributed among the 15 experts in the field. For the calculation of the CVR, the responses were divided into three categories as follows; "Necessary", "It is useful, but not necessary" and "Not necessary". The CVR was calculated based on the questionnaires filled in by the experts:

$$CVR = \frac{ne - \frac{N}{2}}{\frac{N}{2}}$$

ne: The number of experts who have chosen the "necessary" option

N:  Total number of experts

After calculating the CVR, we calculated the Content Validity Index (CVI) to ensure that the best questions were selected for the questionnaire. The "Simplicity", the "Transparency" and the "Relevance" of the CVI were rated by the experts on a Likert scale for each question. Finally, the CVI score for each element was determined by adding the highest score that was agreed upon (3 and 4) divided by the number of experts. Subsequently, the S-CVI/AVE was calculated for each question based on the CVI scores.

## 2.9. Cognitive Interview with students

Based on calculation of CVR and CVI, we prepared the new version of the questionnaire. In this step for ensure more clarity of the questions, we used another method, namely the recorded cognitive interview, which is a common approach in the development and validation phase of the questionnaires, especially when we want to obtain detailed information on the understanding and interpretation of the questions by the participants [25].

This step is a crucial step to ensure that questions are clear, consistent, and accurately measure what you intend to study. The recorded cognitive interview can help identify any problems or ambiguities in the questionnaire questions and make appropriate further modifications to improve their quality.

In this step, 5 students were selected and the last version of the questionnaire was administered. These participants had experiences relevant to the context of the study. During the interview, participants thought aloud as they answered questions on the questionnaire. The entire session was recorded including participant responses and comments about their perception and comprehension about the content and meaning of the questions. Then we analyzed the recordings to identify any problems in understanding the questions, ambiguities. Based on the analyzes of the recorded interviews, the necessary modifications were made to the questionnaire to make the questions clearer and more understandable. But in our case only minor changes have been made.

## Results:

In this section the results of the total procedure which is described above is presented. This section ends to a final version of the I-SCS. It is the Italian version of the System Causability Scale.

After calculating the CVR, the score of question number 8 was not suitable at all, and therefore that question was removed totally. For the other 3 questions (including items number 1,3 and 5) with a score slightly lower than the minimum threshold, after making changes to their wording, with the aim of increasing simplicity and transparency these questions remained in the final version of the questionnaire. So the total number of the questions was 9 which is presented below (table1).

Based on the total number of experts who participated in the CVR section, the minimum accepatance level of each question was 0.49. So each question with less than this level was removed. In this section only the question number 8 is removed. In the CVI section, if a question get a score above 0.6, we remined it and tried to improve its simplicity and transparency.

**Table 1:** The final version of the I-SCS

| Number | Item | CVR | CVI | Final result |
|---|---|---|---|---|
| 1* | Ho trovato che i dati includevano tutti i fattori causali noti rilevanti, con un livello di precisione e dettaglio adeguato | 1 | 0.62 | Remianed with some improvements |
| 2 | Ho compreso le spiegazioni nel contesto del mio lavoro | 0.88 | 0.77 | remained |
| 3* | Sono stato in grado di adeguare il livello di dettaglio in base alle richieste ricevute. | 1 | 0.60 | Remianed with some improvements |
| 4 | Non ho avuto bisogno di supporto per capire le spiegazioni | 0.88 | 0.78 | remained |
| 5* | Ho constatato che le spiegazioni ricevute mi hanno aiutato a comprendere meglio le relazioni di causa-effetto | 0.55 | 0.68 | Remianed with some improvements |
| 6 | Sono stato in grado di utilizzare le spiegazioni grazie alle mie conoscenze di base | 0.88 | 0.71 | remained |
| 7 | Non ho trovato incongruenze tra le spiegazioni | 0.88 | 0.65 | remianed |
| 8** | Penso che la maggior parte delle persone imparerebbe a capire le spiegazioni molto velocemente | 0.08 | 0.38 | Removed from the I-SCS |
| 9 | Non ho avuto bisogno di ulteriori riferimenti nelle spiegazioni: ad es, linee guida mediche, regolamenti | 0.59 | 0.64 | remianed |
| 10 | Ho ricevuto spiegazioni in modo tempestivo ed efficiente | 1 | 0.78 | remianed |

*remained items regards their improvements
**removed from the last version

Based on a proposed method in the original version of SCS, we suggested the 5-point likert type for calculating the last final score of the I-SCS. In this regards "1" means stronglely disagree, "2" means disagree, "3" means neutral, "4" means agree and "5" means stronglly agree. For the calculation of the final score, combination of the total ranking divided by 45 is proposed.

**Discussion:**

The aim of this study was to develope the Italian version of the system causability scale which can be used by xAI developers to assess the quality of explanation of their product regards to the opinion of the end users. So we developed a simple way to assess the usability of a xAI algoritm from the prespective of Italian end-users. The main SCS is inspired by SUS. The latter one is a well-accepted tool for assessing the usability of a product [26]. SUS is used in various studies including Human-computer interaction [27, 28].

SCS is designed to assess the quality of explanations in a xAI data/algoritm. The quality of a explanation is a well-studied concerns in this field [29]. It means that, end-user can use a specific data which is produced based on a AI and satisfy with it (know the reason which produces these results).

Alabi et al. Used this questionnaire for assess the quality of explanation of a a Machine Learning Web-Based Tool for Oral Tongue Cancer Prognostication [30]. They asked 11 pathologists to work with a machine learning-based prognostic tool and then fill the SUS and SCS questionnaire. In another study, Shin et al. Surveyed the role of explainability and casuability on perception, trust and acceptance of a AI-based algorithm. They used SCS for completing their methodology [31]. In the same manner, also other researchers specially from medical fields also reported the quality of explanations of AI-based solutions using SCS tool [32].

Cultural compatibility of measuring tools, especially when the end users are the target of using them, is always one of the important points in validating the tools. This validation and standardization of tools is also found abundantly in human computer interaction and usability studies. For the SUS, there are too many studies can be found regards to cultural adoption and validation of it. Indonesian adaptation and European

Portuguese validation of SUS are reported in two distinct studies [33, 34]. Also other researchers from 30 different countries validated and used it in their own culture [35-37].

Also regards to SCS, we presented the first study to adopted it to Italian context and can be used by Italian end-users.

The limitations of the I-SCS, as like as the original SCS is its likert-type scale. Although this method is different from the scoring method of SUS, but we considered it based on SCS.

The second limitation is it's reliability in time intervals. So we believed that using this tool must be with some cautions. Then we suggest that for more reliable I-SCS, other Italian researchers should design new studies in their field to assess the reliability of this newly developed version.

**Conclusion:**

The outcome of this study was the development of the Italian version of the SCS (I-SCS). This tool can be utilized by Italian researchers and within the Italian cultural context to assess the usability and quality of an xAI system from the perspective of end users.